\documentclass[]{aa}

\def\lcslash#1{/\!\!\!#1}
\def\ucslash#1{/\!\!\!\!#1}

\usepackage{graphics}

\begin{document}

\thesaurus{12(02.04.1; 02.14.1; 08.19.4)}

\title{The effective electron mass in core-collapse supernovae}

\author{Stephen J.\,Hardy}

\institute{Max-Plank-Institut f\"ur Astrophysik,
Karl-Schwarzschild-Str. 1, 85740 Garching bei M\"unchen, GERMANY}

\date{Received <date> / Accepted <date>}

\maketitle

\begin{abstract}
Finite temperature field theory is used to calculate the correction to
the mass of the electron in plasma with finite temperature and
arbitrary chemical potential, and the results are applied to the core
regions of type II supernovae (SNe). It is shown that the effective
electron mass varies between 1 MeV at the edge of the SN
core up to 11 MeV near the center. This changed electron mass affects
the rates of the electroweak processes which involve electrons and
positrons. Due to the high electron chemical potential, the total
emissivities and absorptivities of interactions involving electrons
are only reduced a fraction of a percent. However, for interactions
involving positrons, the emissivities and absorptivities are reduced
by up to 20 percent. This is of particular significance for the
reaction $\bar{\nu} + p \leftrightarrow e^+ + n$ which is a source of
opacity for antineutrinos in the cores of type II SNe. 

\end{abstract}

\section{Introduction}
The properties of electrons are modified by the medium through which
they propagate. In particular, the energy-momentum relation for
electrons is changed from the well-known $E^2= p^2 + m_e^2$ to a
different relation. This change may be modeled by defining a momentum
dependent effective electron mass, $m_{\rm eff}^2 = E^2 - p^2$, which
replaces the vacuum mass of the electron in all calculations involving
electrons and positrons. This has important astrophysical
consequences.

The consequences of the effective mass of the electron has been the
subject of much interest recently in the context of Big Bang
Nucleosynthesis (BBN) (e.g. Chapman \cite{chapman}). During BBN,
plasma effects at temperatures of around 1 MeV lead to a small change
in the effective electron mass, leading to modified neutron-proton
transition rates. This in turn leads to a small change (generally less
than 1 percent) in the abundance of light elements produced by BBN
(Dicus \cite{dicus}; Sawyer \cite{sawyer}). It is possible that
consequences of such small changes in the early universe will have
observable consequences with the advent of the next generation of
cosmic microwave background satellites (Lopez and Turner \cite{lopez}).

On the other hand, the implications of the change in electron mass at
the higher temperatures and densities found in the cores of type II
SNe remain unexplored. Typically, finite temperature
calculations of relevance to BBN are carried out at zero chemical
potential, as there are essentially similar numbers of electrons and
positrons in the early universe. In the highly degenerate cores of
core-collapse SNe this assumption breaks down, and the finite chemical
potential effects must be included. The extension of existing theory
to this regime is made here, where it is shown that the effective mass
of the electron increases with increasing chemical potential, and that
for the densities of relevance in the cores of type II SNe, the
electron mass may be over 20 times its vacuum mass.

One might assume that an increase in the effective mass of the
electron of this magnitude must have immediate drastic consequences
for the physics of type II SNe, as the electron takes part in various
processes important to the neutrino opacity in the core. This is not
the case. While electrons play a significant role in the transport of
neutrinos through the core of a SNe, generally the electron chemical
potential is so high that Fermi blocking ensures that the
electrons involved in neutrino reactions are ultra-relativistic. In
this regime, the finite mass of the electron leads to a small
correction to the rates in the ultra-relativistic limit. Increasing the
electron mass increases this correction, but it remains relatively
small. 

On the other hand, the mass of the positron is also changed, and is
essentially equal to the mass of the electron. As the positron
reactions are not affected by Fermi blocking, the interaction rates
involving positrons may be substantially modified. It is shown here
that these corrections in the cores of SNe are of the order of 10 to 20
percent.

The basic interaction rates between neutrinos and nucleons, nuclei,
electrons, and positrons have been known for many years (see the
Appendices of Bruenn (\cite{bruenn}) for a summary). More recently,
the changes in these rates due to many body effects in the dense
nuclear medium have been examined (e.g. Reddy et
al. \cite{reddy}). This has lead to a decrease in the total neutrino
opacities, though the exact size of this decrease is still the subject
of some theoretical uncertainty (Raffelt \& Seckel
\cite{raffelt}). Here, a different effect is considered, that of
finite temperature effects on the electrons in the medium.

A complete calculation of the lowest order finite temperature
contributions to the neutrino absorption, emission and scattering
rates in a SN core requires the inclusion of not just the modified
electron mass, but also absorption of particles from the plasma,
stimulated emission (which is Fermi blocking for fermions), and the
use of finite temperature wave functions (Donoghue \& Holstein
\cite{donoghue}). This last point is problematic, as there are
technical difficulties in precisely describing the wave functions for
massive particles in a medium (Chapman \cite{chapman}), especially for
the regime considered here, where the electron mass is many times its
vacuum mass. There are essentially three prescriptions for the
construction of finite temperature wave-functions in the literature
(Donoghue \& Holstein \cite{donoghue}; Sawyer \cite{sawyer}; Esposito
et al. \cite{esposito}), and there is some confusion as to which
prescription is correct. Given this difficulty, this paper is
restricted to a calculation of the effects due to the increased
electron mass and Fermi blocking. Preliminary estimates based on the
work of Sawyer (\cite{sawyer}) suggest that these are the dominant
effects.

The plan of this paper is as follows. In the following Section, the
theory for calculating the electron mass at finite temperature and
chemical potential is presented, then these forms are evaluated for
the conditions expected within a core collapse SN. In
Sec.~\ref{sec:Reaction-rates-1}, the modifications to both
$\beta$-decay type reactions, and neutrino scattering off electrons
and positrons are calculated. Finally, the conclusions of this work
are drawn.

\section{Electron mass} \label{sec:Electron-mass}
It has long been recognized that photons propagating in a plasma have
behaviour quite different to those propagating in vacuo. These changes
include a modified dispersion relation (due to a finite plasma
frequency), the appearance of new modes, different polarization
properties, and some momenta for which the photons cannot
propagate. In fact, such effects apply to all particles, though
the magnitude of the changes induced by a medium are dependent on the
strength of the interactions between the particle and the medium. For
instance, in the cores of type II supernovae, neutrons and protons
(SNe) have effective masses different from their vacuum masses
(Horowitz \& Serot \cite{horowitz}). Also, it is known that neutrinos
acquire an induced mass in a medium allowing them to undergo resonant
mixing (the MSW effect) (Mikheyev \& Smirnov
\cite{mikheyev}). Similarly, the properties of electrons are modified
by the medium through which they propagate, and this leads to a
changed dispersion relation (i.e. an effective mass which is momentum
dependent), and changed polarization properties (Weldon \cite{weldon}).

\begin{figure}
\begin{center}
\resizebox{5cm}{!}{\includegraphics{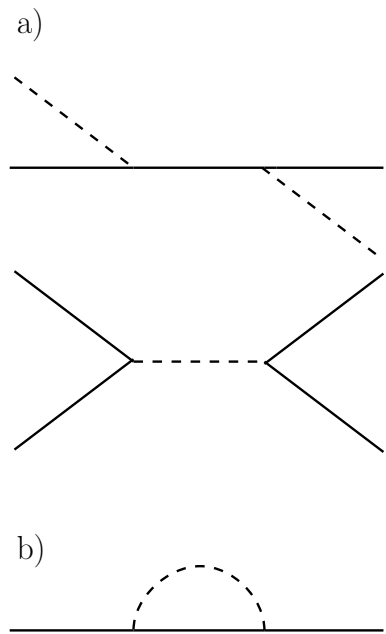}}
\end{center}
\caption{Lowest order Feynman diagrams representing the interactions
which lead to the additional self-energy and hence the additional mass
of the electron in a plasma.}
\label{hardy.fig0}
\end{figure}

The effective mass of the electron at finite temperature may be
understood as an application of the optical theorem to
electrons: an electron in a plasma acquires an energy momentum
relationship different to that in vacuo due to forward scattering
interactions with the photons and electrons in the plasma. Changing to
the notation of Weldon (\cite{weldon}), if the
electron energy is $\omega$ and its momentum is $k$, the effective
mass of the electron is then given by
\begin{equation}
m_{\rm eff}^2 = \omega^2 - k^2.
\end{equation}
To lowest order in the fine structure constant, $\alpha$, the
reactions that contribute to the effective mass of the electron are
shown in Fig.~\ref{hardy.fig0}a. These are just the diagrams for
Compton scattering and Moller scattering. The physical content of this
idea may be expressed more rigorously in the language of finite
temperature field theory. In this theory, the correction of the
electron mass comes about through finite temperature corrections to
the electron self-energy diagram, shown in Fig.~\ref{hardy.fig0}b.

A general procedure for calculating the one-loop correction to the
electron mass was developed by Weldon (\cite{weldon}). Weldon
calculated the effective mass of massless fermions at
finite temperature in the limit of high temperature with zero chemical
potential for a variety of field theories. The procedure applicable to QED with
finite electron mass was also outlined, and this method was applied
by Donoghue and Holstein (\cite{donoghue}), who gave approximate
expressions for the electron mass at finite temperature and zero
chemical potential. This expression was further approximated by Ahmed
and Saleem (\cite{ahmed87a}), who also extended their approximation to
finite chemical potential (Ahmed \& Saleem \cite{ahmed87b}).

The procedure outlined by Weldon (\cite{weldon}) is followed here,
allowing for arbitrary chemical potential. Initially, no approximation
such as that used by Donoghue and Holstein (\cite{donoghue}) are made,
and this leads to some interesting observations concerning the nature
of the collective modes of the electrons at low energy, such as the
existence of a number of additional propagating modes at low momenta.

The correction to the mass of the electron may be calculated from the
electron self-energy diagram shown in Fig.~\ref{hardy.fig0}. In general, the
self-energy, $\Sigma$, at finite temperature will be a linear
combination of $K$, the 4-momentum of the electron, $u$, the
4-velocity of the background plasma, and $1$, the unit matrix. Thus,
the self energy may be written,
\begin{equation}
\Sigma = -a \ucslash{K} -b \lcslash{u} -c', \label{eq:1}
\end{equation}
where $a$, $b$, and $c'$ are functions of the Lorentz invariants
\begin{equation}
\omega = (K \cdot u), \qquad k = \left[(K \cdot u)^2 - K^2\right]^{1/2},
\end{equation}
which, in the frame of the background plasma, may be identified as the
energy and momentum of the electron.

Now, including the correction for the self-energy, the electron
propagator may be written,
\begin{equation}
S^{-1} = \ucslash{K} - m - \Sigma = (1+a) \ucslash{K} + b \lcslash{u} - c,
\end{equation}
with $c=-c'+m$. As discussed in Weldon (\cite{weldon}), the energy-momentum
relation for the propagating electron modes of the system are given by
the location of the poles of the propagator. These poles are
determined by the zeroes of the function
\begin{equation}
D(k,\omega) = (1+a)^2 K^2 + 2(1+a)b (K\cdot u) + b^2 - c^2 = 0, \label{eq:3}
\end{equation}
and the propagator is given by
\begin{equation}
S = [(1+a)\ucslash{K} + b \lcslash{u} + c]/D.
\end{equation}

If one defines
\begin{equation}
A = -{1\over 4} {\rm Tr}[\ucslash{K} \Sigma], \quad B = -{1\over 4} {\rm
Tr}[\lcslash{u} \Sigma], \quad C = -{1\over 4} {\rm Tr}[\Sigma],
\label{eq:14}
\end{equation}
then using Eq.~(\ref{eq:1}), one obtains,
\begin{eqnarray}
a & = & {1 \over k^2} (B \omega - A), \nonumber\\
b & = & {(k^2 - \omega^2) B + \omega A \over k^2}, \nonumber\\
c' & = & C. \label{eq:6}
\end{eqnarray}
The use of the one loop expression for $\Sigma$ in evaluating $A$,
$B$, and $C$, then defines the constants in dispersion function,
Eq.~(\ref{eq:3}). The zeroes of this function may then be found,
determining the electron energy-momentum relation and hence
the effective mass of the electron.

To determine the above coefficients it is necessary to explicitly
calculate the finite-temperature contribution to the electron
self-energy. The algebraic form corresponding to the diagram shown in
Fig.~\ref{hardy.fig0}b may be written
\begin{equation}
\Sigma(K) = i e^2 \int {d^4 p \over (2 \pi)^4} D_{\mu\nu}(p)
\gamma^\mu S(p+K) \gamma^\nu,\label{eq:2}
\end{equation}
where the propagators used are finite temperature propagators given by
\begin{equation}
D_{\mu\nu}(p) = -g_{\mu\nu}\left[ {1 \over p^2 + i0} - 2\pi i
\delta(p^2)n_b(p)\right],
\end{equation}
for the bosons, with $n_b(p)$ the boson occupation number, and 
\begin{equation}
S(p) = (\lcslash{p} + m) \left[ {1 \over p^2 - m^2 + i0} + i {N(p)\over
2 m} \right],
\end{equation}
with $N(p) = \sum_{\epsilon = \pm} 4 \pi m \delta(p^2 - m^2)
H(\epsilon p^0)n^\epsilon(p)$, where $H$ denotes the Heaviside step
function, and $n^+(p)$ and $n^-(p)$ are the electron and positron
distribution functions, respectively. (Note $+$ denotes the particle,
and $-$ denotes the antiparticle.)

One may shift the integration in the fermion term of
Eq.~(\ref{eq:2}) through $p\rightarrow p-K$, leading to
\begin{eqnarray}
\Sigma(K) & = & 2 e^2 \int {d^4 p \over (2\pi)^4} \left[{(2m -
\lcslash{p}) N(p) \over 2m((p-K)^2 + i0)} \right. \nonumber \\ & &
\left. \qquad - {(2m - \lcslash{p} - \ucslash{K}) 2\pi \delta(p^2)
n_b(p) \over (p+K)^2 - m^2 + i0} \right], \label{eq:4}
\end{eqnarray}
where only term involving a single distribution
function has been kept. (The term involving no distribution functions
is the vacuum self energy of the electron, which is removed
by renormalization, and the term involving two distribution functions
is acausal, and thus not physical.) Writing
\begin{eqnarray}
L_1 & = & {\rm ln} \left[ {2(\omega \varepsilon + k p) - K^2 - m^2 \over
2(\omega \varepsilon - k p) - K^2 - m^2} \right], \nonumber \\
L_2 & = & {\rm ln} \left[ {2(\omega \varepsilon + k p) + K^2 + m^2 \over
2(\omega \varepsilon - k p) + K^2 + m^2} \right], \nonumber \\
L_3 & = & {\rm ln} \left[ {2 p (\omega + k) + K^2 - m^2 \over
2 p (\omega - k) + K^2 - m^2} \right], \nonumber \\
L_4 & = & {\rm ln} \left[ {2 p (\omega + k) - K^2 + m^2 \over
2 p (\omega - k) - K^2 + m^2} \right],
\end{eqnarray}
with $\varepsilon = (p^2 + m^2)^{1/2}$, one may perform three of
the integrals in Eq.~(\ref{eq:4}), leading to
\begin{eqnarray}
A_b & = & - {e^2 \over 16 \pi^2 k} \int_0^\infty dp\, n_b(p) \left[8 k p
\right. \nonumber \\ & & \qquad \left. + (K^2 + m^2)(L_3 - L_4)\right] \nonumber \\
A_f & = & - {e^2 \over 16 \pi^2 k} \int_0^\infty {dp\, p \over
\varepsilon} \left[4 k p (n^+(p) + n^-(p))\right. \nonumber \\ & &
\qquad \left. + (K^2 + m^2)(n^+(p) L_1 -
n^-(p) L_2)\right] \nonumber \\
B_b & = & - {e^2 \over 8 \pi^2 k} \int_0^\infty dp\, n_b(p)
\left[(p+\omega)L_3 - (p-\omega)L_4 \right] \nonumber \\
B_f & = & - {e^2 \over 8 \pi^2 k} \int_0^\infty dp\, p 
\left[n^+(p) L_1 + n^-(p) L_2 \right] \nonumber \\
C_b & = & {m e^2 \over 4 \pi^2 k} \int_0^\infty dp\, n_b(p)
\left[L_3 - L_4 \right] \nonumber \\
C_f & = & {m e^2 \over 4 \pi^2 k} \int_0^\infty {dp\, p \over \varepsilon} 
\left[n^+(p) L_1 - n^-(p) L_2 \right], \label{eq:5}
\end{eqnarray}
where the constants of Eq.~(\ref{eq:14}) have been separated into
boson and fermion contributions, according to the appearance or the
relevant distribution function from Eq.~(\ref{eq:4}).
Eq.~(\ref{eq:5}) reduces to the expressions given by Weldon
(\cite{weldon}) in the limit of zero electron mass and zero chemical
potential (i.e. $n^+ = n^- = n_f$). In the limit of zero chemical
potential and making the same approximations, Eq.~(\ref{eq:5}) reduces
to that of Donoghue and Holstein (\cite{donoghue}). Both of these
results also reduce to that of Peressutti and Skagerstam
(\cite{peressutti}) in the limit of zero electron momentum.

To determine the electron mass at a given momentum, the electron
dispersion relation, Eq.~(\ref{eq:3}), must be solved numerically
using Eq.~(\ref{eq:5}). Rewriting Eq.~(\ref{eq:3}) in terms
of $A$, $B$, and $C$ through Eq.~(\ref{eq:6}) leads to
\begin{equation}
A^2 - 2 A(k^2+B \omega) + B^2 K^2 + k^2((C-m)^2 - K^2) = 0, \label{eq:7}
\end{equation}
with $A=A_f + A_b$, etc.

Integration over the logarithmic functions in Eq.~(\ref{eq:5}) is
complicated by the singularities caused by the zeros in the numerators
and denominators. These zeros are located at
\begin{equation}
p_{1\pm} = {(\omega \pm k)^2 \mp m^2 \over 2(\omega \pm k)},
\end{equation}
for $L_1$ and $L_2$, and at $-p_{2\pm}$ for $L_3$ and
$p_{2\pm}$ for $L_4$, with
\begin{equation}
p_{2\pm} = {\omega^2 - k^2 - m^2 \over 2 (\omega \pm k)}.
\end{equation}
In general, the coefficients $A$, $B$, and $C$ will be complex valued,
as the arguments of the logarithms in Eq.~(\ref{eq:5}) are
negative for values of $p$ between the two zeros. The logarithms in
Eq.~(\ref{eq:5}) are to be taken in the principle value sense,
and the imaginary part of the coefficients are neglected.

The approximations made by Donoghue \& Holstein (\cite{donoghue}) lead
to the relation
\begin{equation}
m_{\rm eff}^2 = m^2 - 2(A + m C) \label{eq:13}
\end{equation}
which is obviously much simpler to evaluate than the solutions to
Eq.~(\ref{eq:7}), and, as is shown below, is quite accurate.

\subsection{Reaction rates} \label{sec:Reaction-rates}
To calculate reaction rates at finite temperature, it is necessary to
use finite temperature spinors, which are the solutions of a modified
Dirac equation, given by
\begin{equation}
(\ucslash{K} - m - \Sigma) U = 0,\label{eq:11}
\end{equation}
where U is the spinor wave function in the medium. For a massless
electron, it was shown by Weldon (\cite{weldon}) that the wave functions in the
medium are unchanged in form from their vacuum counterparts -- only
their normalization is modified, and the normalization coefficient
may be calculated by determining the residue of the electron
propagator at the electron pole. Donoghue and Holstein (\cite{donoghue})
considered finite temperature effects on the mass of a massive
electron. In this case, in general, the solutions to
Eq.~(\ref{eq:11}) do not correspond to the vacuum
solutions. Donoghue and Holstein made an assumption that a
corresponding set of states did exist and calculated the normalization
coefficient based on this assumption. The coefficient is given by
\begin{equation}
Z_2^{-1} = 1 - \left. (\omega + B)^{-1}\left(\omega + {\partial A \over
\partial \omega} + m {\partial C \over \partial \omega}
\right)\right|_{\omega=\sqrt{k^2 + m^2}}. \label{eq:12}
\end{equation}
Sawyer (\cite{sawyer}) reexamined this procedure and made a
perturbation series expansion around the vacuum poles of the
propagator, effectively calculating the projection operators to be
used within QED at finite temperature. Esposito et
al. (\cite{esposito}) also adopted a peturbative approach, but obtain
a result different from both Sawyer and Donoghue and Holstein.

For this work, a further problem presents itself. As is shown below,
electrons in the core of a SN may have masses many times their vacuum
mass. Thus, it is unclear whether a perturbation series expansion
around the vacuum poles of the propagator is a valid approach to this
problem at all. In the absence of a well defined theory, it is assumed
that wave-function renormalization effects are unimportant in
comparison to the kinematical effects of the changed electron mass. It
is unlikely that the correct inclusion of normalization effects will
have a significant effect on the results presented, as, in the regime
of interest, the high electron chemical potentials ensure that the
electrons are relativistic. This implies that the vacuum mass of the
electron is relatively unimportant, rendering the massless approach of
Weldon (\cite{weldon}), in which there are no normalization effects,
approximately valid.

\section{Evaluation of the effective mass} \label{sec:Eval-effect-mass}

\begin{figure}
\resizebox{\hsize}{!}{\includegraphics{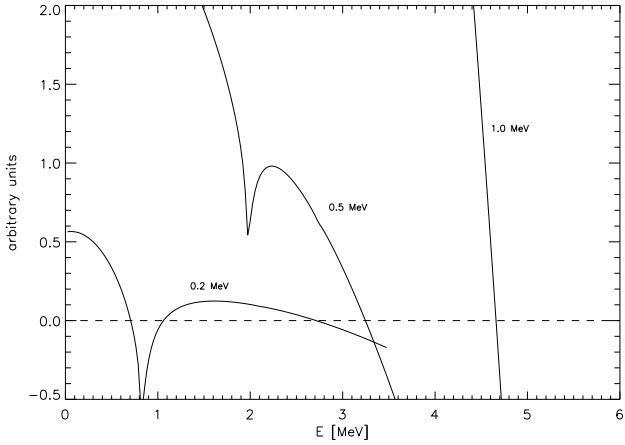}}
\caption{Dispersion functions for electrons in a plasma with
temperature of 1 MeV and chemical potential of 10 MeV, for a variety
of electron momenta. The physical energy of the electron is
given by the value of the abscissa for which the dispersion function
is zero. Note that multiple roots appear at low momenta.}
\label{hardy.fig1}
\end{figure}

To evaluate the effective mass of the electron, the locations of the
zeroes of the dispersion function, Eq.~(\ref{eq:7}), must be
found numerically. Assuming a plasma temperature of 1 MeV and a
chemical potential of 10 MeV, the dispersion function as a function of
particle energy is shown for a variety of electron momenta in
Fig.~\ref{hardy.fig1}. The point where these dispersion relations pass
through the zero point identifies the energy of the particle for its
given momentum.

The structure of the dispersion functions in Fig.~\ref{hardy.fig1} is
quite complicated, and shows that for low momenta there exist multiple
roots to the electron dispersion equation. An effect analogous to this
has been noted before by Klimov (\cite{klimov}) and Weldon
(\cite{weldon}, \cite{weldon89}), where two solutions to the electron
dispersion relation at zero chemical potential and zero electron mass
were found. These were identified as one root corresponding to the
electron, and one to a propagating collective mode later labeled the
``plasmino'' by Braaten (1992). The second of the lower two roots at
small momenta in Fig.~\ref{hardy.fig1} may be identified as the
plasmino mode. The existence of the third zero in the dispersion
equation of electrons at low momentum has not been noted before, due
to the analytic approximations made in calculating the mode structure.
Whether this zero corresponds to a propagating mode depends on the
ratio of the imaginary part of the electron self-energy to the real
part. If this ratio is comparable to or larger than unity then this
mode is strongly damped and does not propagate.

While the structure of the electron dispersion relation, and the
nature of these new modes are a fascinating subject in their own
right, no analysis of the effects of these new modes will be made
here. Given that they only appear at low momenta, they are unlikely to
have a significant dynamical effect, as the relevant energy scales for
electrons are the temperature and chemical potential, which are
high. Attention is restricted to the electron-like pole of the
electron dispersion relation, avoiding the difficult task of
calculating the polarization and damping characteristics of these
new modes.  Note, however, that these new modes are of interest and should
be investigated.

\begin{figure}
\resizebox{\hsize}{!}{\includegraphics{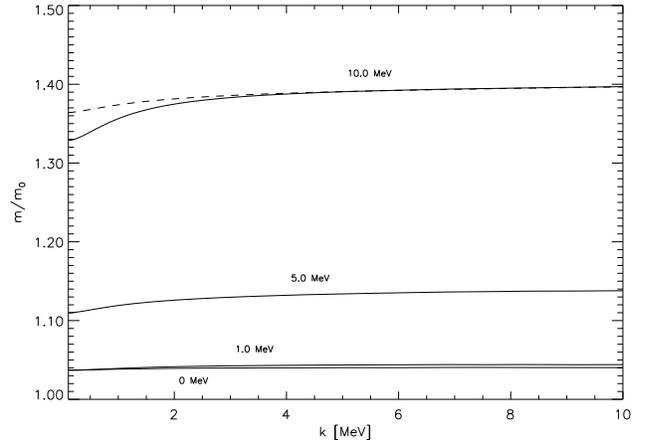}}
\caption{Ratio of the effective electron mass to the vacuum electron
mass as a function of electron momentum for various chemical
potentials at a temperature of 1 MeV. The dashed line corresponds to
the approximation of Donoghue and Holstein (1983).}
\label{hardy.fig2}
\end{figure}

For a variety of values of chemical potential, the effective mass as a
function of momentum is shown in Fig.~\ref{hardy.fig2}. Also shown
is the approximate form of the electron mass as given by
Eq.~(\ref{eq:13}). Clearly, this form is quite a good
approximation to the electron-like pole of the dispersion function,
and may be used for efficient evaluation of the effective electron
mass at all chemical potentials.

\begin{figure}
\resizebox{\hsize}{!}{\includegraphics{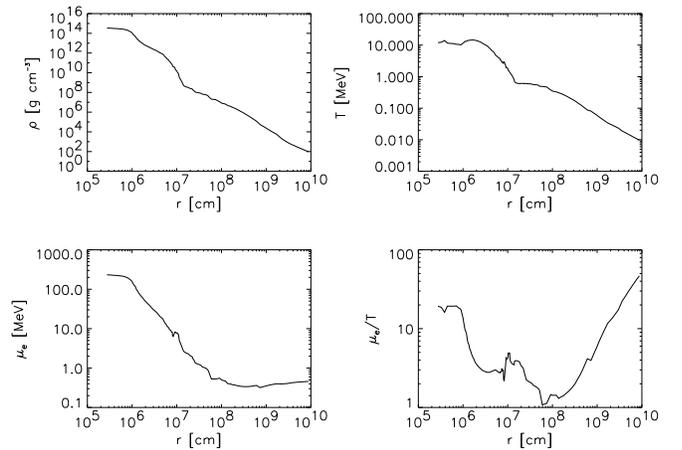}}
\caption{Data from a numerical simulation of a SN by Bruenn.}
\label{hardy.fig4}
\end{figure}

\begin{figure}[ht]
\resizebox{\hsize}{!}{\includegraphics{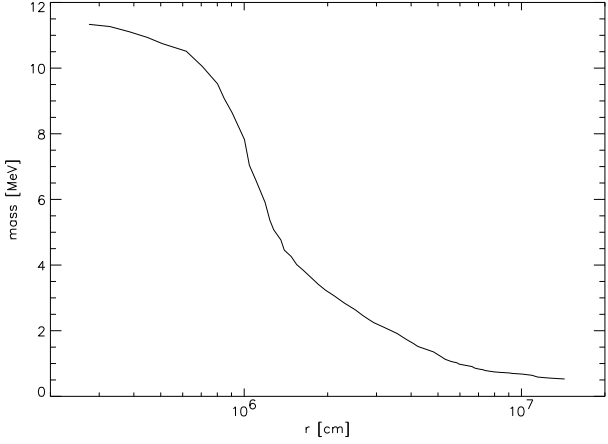}}
\caption{The electron effective mass as a function of distance from
the center of a type II SN.}
\label{hardy.fig5}
\end{figure}

For concreteness, a particular model for the central regions of a
core-collapse SN as generated by S. Bruenn (\cite{bruenn}) is now
considered. The relevant data for this model is shown in
Fig.~\ref{hardy.fig4}. This model is based on a 15 solar mass
progenitor around 0.2 seconds after core bounce. At this stage the
bounce shock has proceeded to around 100 km and the semi-opaque region
for the neutrinos lies between 10 to 50 km. The electron mass as a
function of radius for this model is shown in
Fig.~\ref{hardy.fig5}. This effective mass was calculated for
electrons with their momenta set at half the chemical potential of the
plasma. This ensure the mass is within the flat region of the curves
in Fig.~\ref{hardy.fig2}, and is in the appropriate range for the
reactions considered below.

\subsection{Equilibria}\label{sec:Equilibria}

The increase in the effective mass of the electron in the core regions
of a supernova has an effect on the relative numbers of particles
which form an equilibrium plasma in these dense environments. A
detailed calculation of the equilibrium configurations of an ideal
neutron, proton, electron and neutrino gas at fixed lepton fraction
has been made both with and without the corrections to the electron
mass. The corrections to the quantities characterizing the thermal
equilibria, such as the chemical potentials of the electrons, protons,
neutrons and neutrinos are of the order of $0.1$ percent, and are thus
dynamically unimportant.

\section{Reaction rates}\label{sec:Reaction-rates-1}
The results of Sec.~\ref{sec:Eval-effect-mass} are now applied to
determine the change of neutrino opacities in the cores of SNe.  In
general, it might be argued that the electron mass effects should be
great, as the standard correction to the neutrino scattering rates
scales as $m_e^2/T^2$, which is of order 1 for the region near the
center of the SN. However, for interactions involving degenerate
distributions of electrons, the correction to the rate scales as
$m_e^2/\mu_e^2$, which is always a small quantity in a SN. Thus only
rates of interactions which involve positrons are significantly
modified.

The changes to the absorptivity, $1/ \lambda^{(a)}(\omega)$, for the
processes $\nu_e + n \rightleftharpoons e + p$, $\bar{\nu}_e + p
\rightleftharpoons e^+ + n$, and $\nu_e + A' \rightleftharpoons A + e$
as calculated by Bruenn (\cite{bruenn}) are calculated. The
approximate rates for these processes are given by
\begin{eqnarray}
\nu_e + n & \rightleftharpoons & e + p \nonumber: \\
{1 \over \lambda^{(a)}(\omega)} & = & {G^2 \over \pi} \eta_{np}(g_V^2
+ 3 g_A^2) \left[1 - f_e(\omega+Q)\right] \nonumber \\ & & \qquad \times (\omega+Q)^2 \left[1 - {m_E^2
\over (\omega+Q)^2} \right]^{1 \over 2}; \nonumber \\
\label{eq:8}
\end{eqnarray}
\begin{eqnarray}
\bar{\nu}_e + p &\rightleftharpoons & e^+ + n \nonumber: \\
{1 \over \lambda^{(a)}(\omega)} & = & {G^2 \over \pi} \eta_{pn}(g_V^2
+ 3 g_A^2) \left[1 - f_{e+}(\omega-Q)\right] \nonumber \\
& & \times \, (\omega-Q)^2 \left[1 - {m_E^2
\over (\omega-Q)^2} \right] H(\omega-Q-m_e); \nonumber \\
 \nonumber \\ \,\label{eq:9}
\end{eqnarray}
\begin{eqnarray}
\nu_e + A' & \rightleftharpoons & A + e \nonumber: \\ {1 \over
\lambda^{(a)}(\omega)} & = & {G^2 \over \pi} n_A \exp[\beta(\mu_n -
\mu_p - Q')] g_A^2 \nonumber \\ & & \qquad \times {2 \over 7}
N_p(Z)N_h(N) \left[1-f_e(\omega+Q')\right] \nonumber \\ & & \qquad
\times\, (\omega+Q')^2 \left[1 - {m_E^2
\over (\omega+Q)^2} \right]^{1 \over 2};\label{eq:10}
\end{eqnarray}
where $G$ is the Fermi constant, $\eta_{np}$ and $\eta_{pn}$ are the
effective neutron and proton number densities, respectively,
$Q=1.2935$ MeV is the difference between neutron and protons masses,
$N_p$, and $N_h$ are the number of neutrons and neutron holes in the
$1f_{7/2}$ and $1f_{5/2}$ levels, respectively, $Q'$ is approximately
the difference between the neutron and proton chemical potentials plus
3 MeV, and $n_A$ is the number density of particles with
atomic weight $A$. Note that as absolute magnitudes of these rates are
not needed for this calculation, detailed knowledge of many of these
quantities is unnecessary. For further details, consult Bruenn
(\cite{bruenn}).

Including the changes in the nature of the electron at finite
temperature and chemical potential leads to a number of changes in the
above rates. First, the electron mass in
Eqs.~(\ref{eq:8})-(\ref{eq:10}) must be replaced with the
effective electron mass (at high momentum). Secondly, each rate
obtains a factor from the normalization of the external electron line
given in Eq.~(\ref{eq:12}), which is neglected here. Finally, the
conditions relating to the minimum neutrino energy at which the
process occurs are changed, with Eq.~(\ref{eq:8}) requiring an
extra term limiting the process to reactions where the neutrino energy
is higher than $Q - m_{eff}$.

\begin{figure}
\resizebox{\hsize}{!}{\includegraphics{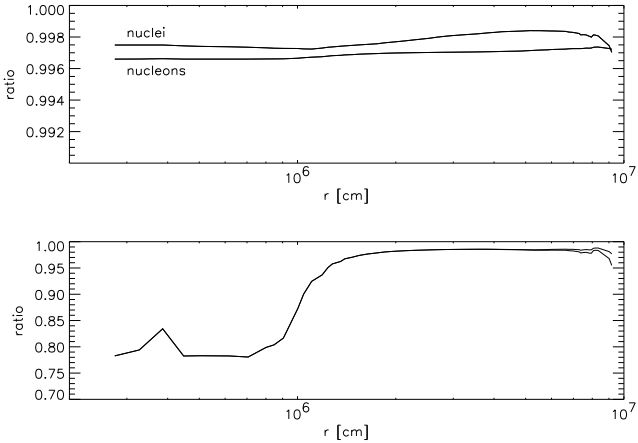}}
\caption{Change in emissivities and absorptivities as a function of
radius for the three nucleon processes mentioned in the text. The top
panel shows that neutrino absorption and emission on nucleons and
nuclei are hardly suppressed at all. The bottom panel shows that in the
high density regions of the proto-neutron star, the absorption and
emission of antineutrinos by nucleons is suppressed by up to 20
percent}
\label{hardy.fig6}
\end{figure}

The ratio of modified to unmodified absorptivities, integrated over a
thermal neutrino spectrum, are shown in Fig.~\ref{hardy.fig6} as a
function of radius from the center of the SN. Note that under local
thermodynamic equilibrium, the absorptivities and emissivities differ
only by a factor of a neutrino distribution function. Thus, for
thermal neutrino spectra, the emissivities and absorptivities will be
changed by the inclusion of the modified electron mass by the same
amount, as the effective electron mass does not directly affect the
neutrino distribution function. The top panel of Fig.~\ref{hardy.fig6}
shows that the neutrino interactions with neutrons and nucleons remain
largely unchanged by the change in the electron mass. This is because
the high electron chemical potentials found in the core of the SN
restrict the energy of the created electron to very high energies. At
these energies, the electron mass is a smaller fraction of the
electron energy, and thus the correction due to the finite mass of the
electron is only a small correction to the rate.

\begin{figure}
\resizebox{\hsize}{!}{\includegraphics{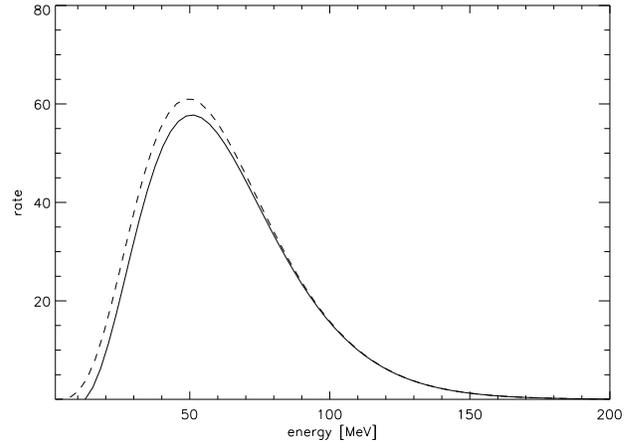}}
\caption{Change in the absorptivity of antineutrinos on protons in the
central regions of the core as a function of energy (arbitrary
vertical scale). The unmodified rate is the dashed line, and the rate
modified by the changed electron mass is shown as the solid line. The
effect of the additional effective mass are largest at low
energies. The average energy of the neutrinos in this region is around
30 MeV.}
\label{hardy.fig7}
\end{figure}

The bottom panel of Fig.~\ref{hardy.fig6} shows that the
absorptivities (and emissivities) of antineutrinos is reduced by a
substantial amount. This is because the positrons are not restricted
to high energies by Fermi blocking. This may be seen in more detail in
Fig.~\ref{hardy.fig7}, where the ratio of the absorptivities is shown as a
function of neutrino energy. Clearly the absorptivity is reduced much
more at low energy (and the cut off for absorption is at a higher energy
in the modified case). This may also be significant in the
region between 10 and 20 km in which the plasma is only partially
opaque to neutrinos and some energy equilibration between the
neutrinos and the stellar plasma occurs. Although the total reduction in
the rates is smaller in this region, the change in the energy
distribution of the antineutrinos may have an effect on the predicted
spectrum of anti-neutrinos observable from a galactic
supernova. 

\subsection{Neutrino-electron scattering}
The calculation of neutrino electron scattering rates including
electron mass corrections is an involved process, and for the
sake of brevity, the scattering kernels will not be reproduced
here. The reader is referred to Schinder and Shapiro (\cite{schinder})
Eq.~(47), which is the form of the rate used here. 

\begin{figure}
\resizebox{\hsize}{!}{\includegraphics{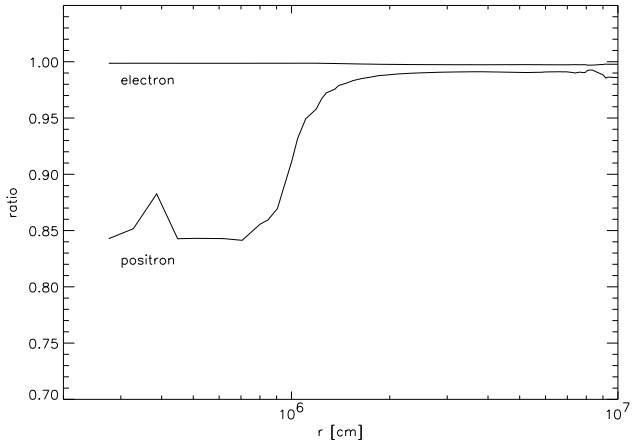}}
\caption{Change in scattering rates as a function of
radius for neutrino scattering off electrons and positrons. The total
neutrino-electron scattering cross-sections are only reduced by around
0.3 percent.}
\label{hardy.fig8}
\end{figure}

In Fig.~(\ref{hardy.fig8}) the ratio of rates with and without a
modified electron mass for neutrino-electron scattering and
neutrino-positron scattering are integrated over ingoing and outgoing
thermal distributions of neutrinos as a function of radius from the
center of a model SN. Again it is shown that the rates involving
electrons are not changed substantially, but interactions with
positrons are reduced by about 15 percent. This is not a dynamically
important effect for models of type II SNe, as the number of positrons
in the core is exponentially suppressed by the very large chemical
potential.

While there is relatively little change in the total opacity of
neutrinos to electrons caused by the increased effective electron
mass, there could be important changes introduced in the neutrino
spectrum due to the increased elasticity of the scatterings. This is
due to the dominant role that neutrino electron scatterings play in
the energy exchange between the neutrinos and the electrons in the
semi--transparent region found at densities of around $10^{13}$ $
\mbox{g cm}^{-3}$.

It is worth noting also that at temperatures of greater than 1 MeV, the
efficiency of electron positron production by neutrinos is decreased
by the increase in the electron mass. This may be of relevance to
Gamma Ray Burst scenarios which rely on neutrino heating. Though this
effect is not studied in detail here, we expect the reduction to only
be a matter of a few percent at 1 MeV (as the chemical potentials are
small compared to the temperature) though it will be significantly
higher at higher temperatures.  Alternately, the rate of energy
deposition due to neutrino electron scattering could be increased due
to the additional mass of the electron at high temperature.

\section{Conclusions}

It has been shown that electrons within the core regions of core
collapse SNe behave as if they have a mass many times their vacuum
mass. In particular, the effective electron mass varies from around 11
MeV in the central regions of the core, to 1 MeV at the edge of the
core. The primary effect of this increased electron mass is the
reduction of the antineutrino emissivity and absorptivity off
protons by around 20 percent. A complete calculation of these rates to
lowest order in the fine structure constant requires a finite
temperature treatment of the radiative corrections to the rate of
$\bar{\nu}_e + p \rightleftharpoons e^+ + n$, including a complete
description of the finite temperature wave function of an electron in
these dense environments.

\section*{Acknowledgments}
The author wishes to acknowledge valuable discussions with
W. Hillebrant, G. Raffelt, and T. Janka, who also supplied the
numerical supernova model used in this work. This work was
carried out with the support of a TMR Marie Curie Fellowship and
the Deutsche Forschungsgemeinschaft project SFB 375-95.

\end{document}